\documentclass{moriond}

\usepackage{amssymb}

\def\Journal#1#2#3#4{{#1} {\bf #2}, #3 (#4)}

\def\NPB{{\em Nucl. Phys.} B}
\def\PLB{{\em Phys. Lett.}  B}

\def\PRD{{\em Phys. Rev.} D}
\def\ZPC{{\em Z. Phys.} C}

\def\be{\begin{equation}}
\def\ee{\end{equation}}
\def\bea{\begin{eqnarray}}
\def\eea{\end{eqnarray}}

\newcommand{\alphas}{\alpha_{\rm s}}
\newcommand{\alphasmZ}{\alphas(\rm m^2_{_{\rm Z}})}
\newcommand{\sqrts}{\sqrt{\rm s}}

\newcommand{\lqcd}{\Lambda_{_{\rm QCD}}}

\newcommand{\epem}{e^+e^-}
\newcommand{\meff}{m{_{\rm eff}}}

\providecommand{\bbbar}{\rm b\overline{b}}

\newcommand{\Photo}{\includegraphics[height=37mm,natwidth=213,natheight=290]{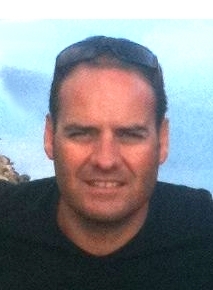}}

\begin{document}
\vspace*{-1.cm}
\title{Extraction of $\alphas$ from the energy evolution of jet fragmentation functions at low $z$}


\author{\underline{David d'Enterria}$^1$, and Redamy P\'erez-Ramos$^2,$\footnote{R.~P\'erez-Ramos 
acknowledges support from the Academy of Finland, Projects No. 130472 and No.133005.}}

\address{$^1$  CERN, PH Department, CH-1211 Geneva 23, Switzerland\\
$^2$ Department of Physics, Univ. of Jyv\"askyl\"a, P.O. Box 35, F-40014 Jyv\"askyl\"a, Finland}

\maketitle\abstracts{
A novel method to extract the QCD coupling $\alphas$ from the energy evolution of the moments of the
parton-to-hadron fragmentation functions at low fractional hadron momentum $z$, is presented. 
The evolution of the moments (multiplicity, peak, width, skewness) of the charged-hadron distribution in jets
is computed at NMLLA+NLO$^*$ accuracy and compared to the experimental 
deep-inelastic $e^\pm$p data. Values of the strong coupling constant at the Z pole are obtained,
$\alphasmZ$~=~0.119$\pm$0.010, in excellent numerical agreement with the current world average
determined using other methods at the same level of accuracy.}

\section{Introduction}

In the chiral limit of massless partons, quantum chromodynamics (QCD) has one single fundamental parameter:
its coupling constant $\alphas$, determined at a given reference energy scale $Q$. Starting from a
value of $\lqcd\approx$~0.2~GeV, where the perturbatively-defined coupling diverges, the value of $\alphas$ 
decreases with increasing energy following a generic $\ln(Q^2/\lqcd^2)$ dependence. The current  $\alphas$ world 
average~\cite{PDG} at the Z mass pole, $\alphasmZ$~=~0.1185$\pm$0.0006, has a $\pm$0.5\%
uncertainty, making of the strong coupling the least precisely known of all fundamental constants in
nature. Reducing the  $\alphas$ uncertainties to the permille level is a prerequisite (i) in calculations of
higher-order corrections to {\it all} (partonic) cross sections at hadron colliders, (ii) for precision fits
of the Standard Model ($\alphas$ dominates e.g. the Higgs boson H$\to\bbbar$ partial width uncertainty), 
and (iii) for searches of physics beyond the SM (e.g. running of the interaction couplings up to the grand
unification scale).

In this context, it is of utmost importance to find new independent approaches to determine $\alphas$ from the data, 
with different experimental and theoretical uncertainties than the methods currently used. 
In Ref.~\cite{NMLLA_NLO} we have presented a new technique to obtain $\alphas$ from the energy evolution of
the moments of the parton-to-hadron fragmentation functions including, for the first time,
higher-order NNLL logarithms resummations and NLO running-coupling corrections. 
The approach, tested with experimental jet data from $\epem$ annihilation, is extended here to cover also
deep-inelastic $e^\pm$p collisions in the current hemisphere of the Breit (or ``brick wall'') frame where the
incoming quark scatters off the photon and returns along the same axis.

\section{Evolution of the parton-to-hadron fragmentation functions at NMLLA+NLO$^*$}

The distribution of hadrons in a jet is theoretically encoded in a 
fragmentation function (FF), $D_{\rm i\to h}(z,Q)$, describing the probability that the parton $i$ fragments
into a hadron $h$ carrying a fraction $z$ 
of the parent parton's momentum. 
Although one cannot compute perturbatively the FFs at a given scale $Q$, 
their evolution, i.e. the process of parton radiation and splitting occurring in a jet shower,
can be theoretically predicted. The evolution of the FF from a scale $Q$ to $Q'$ is governed at large
$z\gtrsim 0.1$ by the DGLAP~\cite{dglap} equations, and at small $z$ by Modified Leading Logarithmic
Approximation (MLLA)~\cite{mlla} approaches which resum soft and collinear singularities. 
The set of integro-differential equations for the FF evolution --including next-to-leading-order $\alphas$ and
next-to-NMLLA corrections-- have been solved in Ref.~\cite{NMLLA_NLO} by expressing the Mellin-transformed hadron
distribution in terms of the anomalous dimension $\gamma$:
$D\simeq C(\alphas(t))\exp\left[\int^t \gamma(\alphas(t')) dt\right]$ for $t=\ln Q$, resulting in an
expansion in (half) powers of $\alphas$: 
$\gamma\sim {\cal O}_{_{\rm DLA}}(\sqrt{\alphas})+{\cal O}_{_{\rm MLLA}}(\alphas)+{\cal O}_{_{\rm
 NMLLA}}(\alphas^{^{3/2}})+\cdots$, of which two new higher-order terms have been computed for the first time.

Writing the FF as a function of the log of the inverse of $z$, $\xi=\ln(1/z)$, emphasizes the region of
relatively low momenta that dominates the spectrum of hadrons inside a jet. 
Due to colour coherence and gluon-radiation interference, not the softest partons but those with
intermediate energies ($E_h\propto E_{\rm jet}^{0.3}$) multiply most effectively in QCD cascades, 
leading to an energy spectrum with a typical ``hump-backed plateau'' (HBP) shape as a function of $\xi$ (Fig.~\ref{fig:1}).
Without any loss of generality, one can express the fragmentation function 
in terms of a distorted Gaussian: 
\begin{equation}
D(\xi,Y,\lambda) = {\cal N}/(\sigma\sqrt{2\pi})\cdot e^{\left[\frac18k-\frac12s\delta-
\frac14(2+k)\delta^2+\frac16s\delta^3+\frac1{24}k\delta^4\right]}\,, \mbox{ with }\delta=(\xi-\bar\xi)/\sigma,
\label{eq:DG}
\end{equation}
and study the evolution of all its moments starting from a parton energy $Y = \ln{E\theta/Q_{_{0}}}$
down to a shower cut-off scale $\lambda = \ln(Q_{_{0}}/\lqcd)$. 
The corresponding expressions for the NMLLA+NLO$^*$ evolutions\footnote{The asterisk in the term 'NLO*' indicates that
there are missing NLO corrections in the  splitting functions.} in $Y,\lambda$ for (i) the charged-hadron 
multiplicity inside a jet N$_{\rm ch}$, 
(ii) the peak position $\bar\xi$, (iii) width $\sigma$, (iv) skewness $s$, and (v) kurtosis $k$, 
have been obtained in Ref.~\cite{NMLLA_NLO}.
If one evolves the fragmentation functions down to the lowest possible scale, $Q_{_{0}} \to \lqcd$
(i.e. $\lambda = 0$, aka. ``limiting spectrum''), one obtains expressions for the HBP moments which depend on a
{\it single} parameter: $\lqcd$ or, equivalently, $\alphas$. Such an approach is justified by the 
``local parton hadron duality'' 
which states that the distribution of partons in jets is identical
to that of the final hadrons (up to a constant ${\cal K}^{\rm ch}$ which only affects the absolute normalization
of the HBP distribution). Thus, by fitting  to the DG parametrization the experimental hadron jet data at
various energies, one can determine $\alphas$ from the corresponding energy-dependence of its fitted
moments.
\section{Data-theory comparison and $\alphas$ extraction}

By applying the methodology described above to the world jet data measured in $\epem$ 
at $\sqrts\approx$~2--200~GeV, we obtained~\cite{NMLLA_NLO} a value of $\alphasmZ$~=~0.1195~$\pm$~0.0022.
\begin{figure}[htpb!]
\centerline{
\includegraphics[width=0.75\linewidth,natwidth=787,natheight=465]{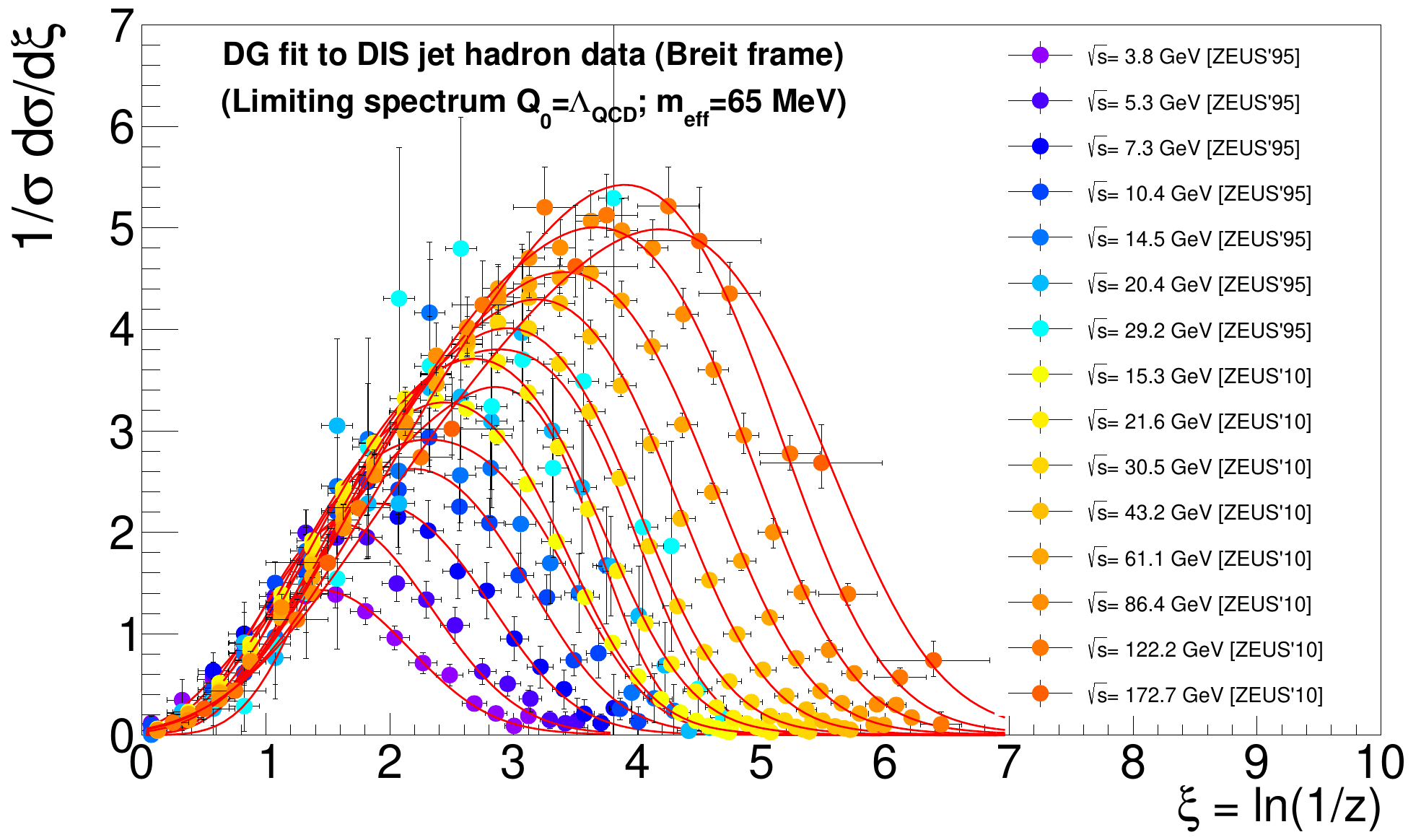}
}
\caption[]{
Charged-hadron distributions in jets measured in $e^\pm$p collisions at
$\sqrts\approx$~4--180~GeV~\cite{ZEUS} as a function of $\xi=\ln(1/z)$
fitted to the distorted Gaussian (DG), Eq.~(\ref{eq:DG}).
}
\label{fig:1}
\end{figure}
In the present work, we confront our NMLLA+NLO$^*$ calculations to all the existing charged-hadron spectra from jets
measured in 
DIS ($e^\pm,\nu$-p) collisions.
In the DIS Breit frame, the kinematic evolution variable equivalent to the $\epem$ squared centre of mass
energy, is the invariant 
four-momentum transfer Q. 
In addition, if one wants to compare $\epem$ and DIS data, the DIS FF have to be scaled up by a factor of two
to account for the fact that they cover the hadronic activity only in half (current) hemisphere of the Breit
frame.
There exist 55 direct measurements of HBP moments (mostly $N_{\rm ch}$ multiplicity, peak, and width)
from H1~\cite{H1old} and ZEUS~\cite{ZEUSold} experiments in $e^\pm$p at HERA, and NOMAD ($\nu$N
scattering)~\cite{Altegoer:1998py} 
covering the range Q~$\approx$~1--180~GeV. Moreover, we have added the moments resulting from the
fitting to Eq.~(\ref{eq:DG}) of the single-hadron distributions (amounting to 250 individual data points) 
measured by ZEUS~\cite{ZEUS} 
and shown in Fig.~\ref{fig:1}. 
Finite hadron-mass effects in the DG fit have been accounted for through a rescaling of
the theoretical (massless) parton momenta with an effective mass $\meff$ as discussed in
Ref.~\cite{NMLLA_NLO}.

\begin{figure}[htpb!]
\centerline{
\includegraphics[width=0.90\linewidth,natwidth=787,natheight=644]{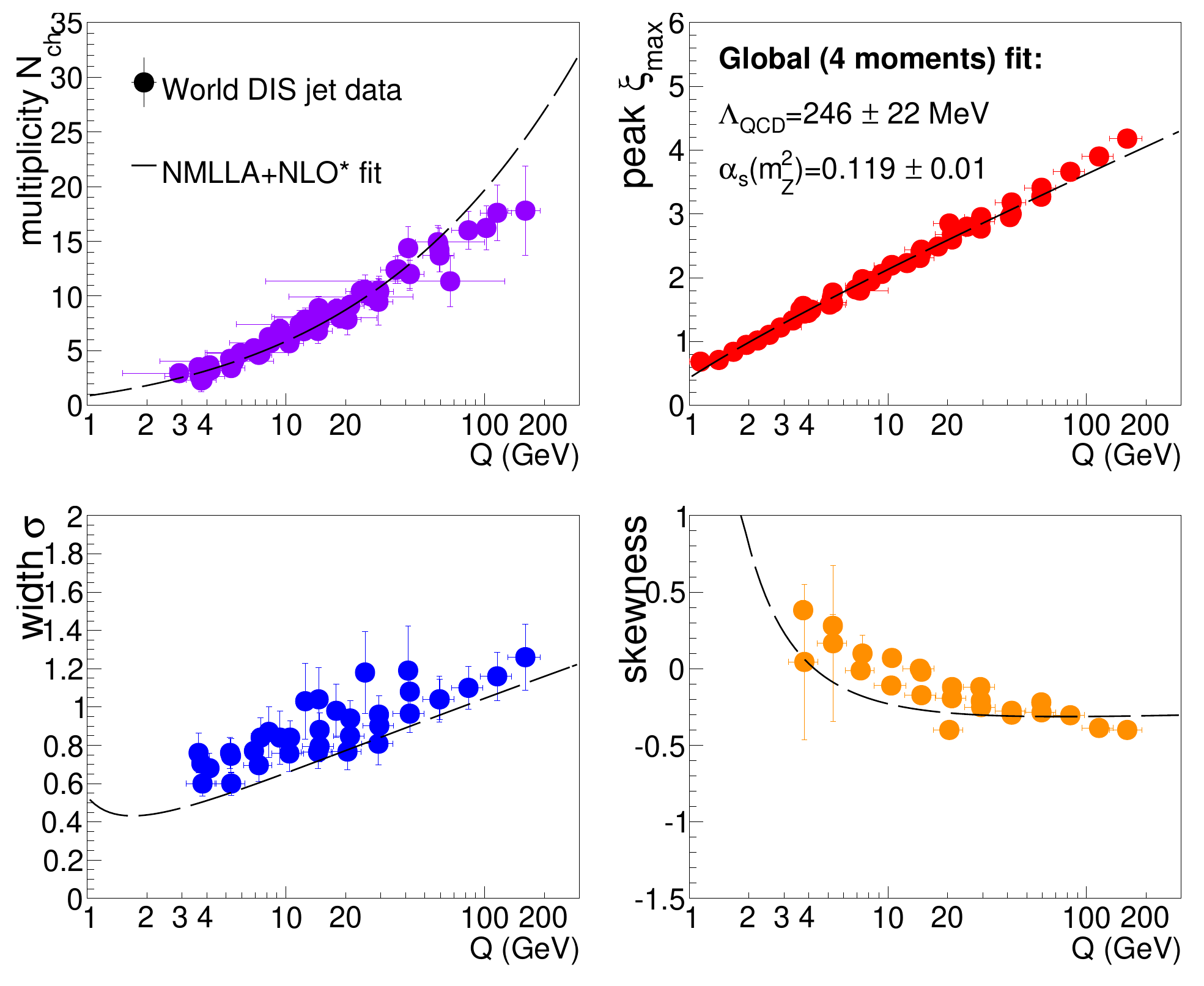}
}
\caption[]{Energy evolution of the moments of the jet FF measured in DIS collisions in the
  range Q~$\approx$~1--180~GeV, fitted to the NMLLA+NLO* predictions for the total charged hadron multiplicity (top,
  left), the peak position (top, right), the width (bottom, left), and the skewness (bottom, right).
  The extracted values of 
  $\lqcd$ and equivalent NLO $\alphasmZ$ 
  are quoted for the combined global fit.} 
\label{fig:2}
\end{figure}

The evolutions as a function of energy of all the extracted FF moments (except the kurtosis which is almost zero
and not properly reproduced by the calculations~\cite{NMLLA_NLO}) are shown in Fig.~\ref{fig:2}.
Hadron multiplicity, peak and width increase with energy, while skewness (and kurtosis, not shown) decrease.
All the energy dependencies are well described by the NMLLA+NLO* predictions for $N_f$~=~5 active quark
flavours\footnote{The moments of the lowest-$\sqrts$ data have a few-percent correction applied to account
for the (slightly) different ($N_f$~=~3,4) evolutions below the charm and bottom production thresholds.}, as shown
by the fitted curves obtained for the limiting spectrum ($\lambda$~=~0) with $\lqcd$ as single free parameter.
[In the case of the total charged-hadron multiplicity, there is an extra free parameter, the overall
normalization constant ${\cal K}^{\rm ch}\approx$~0.10 of the HBP spectrum, which nonetheless plays no role in
the $\lqcd$ determination].
Table~\ref{tab:1} lists the values of the $\lqcd$ parameters (and associated values of $\alphasmZ$ at NLO
accuracy) individually extracted from the energy evolutions of each one of the four DG components, as well as 
the combined global fit (obtained using {\sc minuit2}) of all moments (last column).

\begin{table}[htbp!]
\caption{\label{tab:1} Values of $\lqcd$ and associated $\alphasmZ$ 
  obtained from the fits of the energy-dependence of the moments of the charged hadron distribution of jets
  measured in DIS obtained from their NMLLA+NLO$^{*}$ evolutions. 
  The last column provides the combined value of the strong coupling from a global fit of all FF moments.}  
\begin{center}
\begin{tabular}{lcccc|c}\hline
 DG moment:    & Peak position      & Multiplicity      & Width             & Skewness         & Combined \\\hline
 $\lqcd$ (MeV) &  266 $\pm$ 5$_{\rm (fit)}$      & 178 $\pm$ 37$_{\rm (fit)}$      & 203 $\pm$ 13$_{\rm (fit)}$      & 235 $\pm$ 182$_{\rm (fit)}$    & 246 $\pm$ 22\\
 $\alphasmZ$   &  0.121 $\pm$ 0.007 & 0.114 $\pm$ 0.017 & 0.116 $\pm$ 0.007 & 0.119$\pm$ 0.092 & 0.119 $\pm$ 0.010 \\\hline
\end{tabular}
\end{center}
\end{table}

The most ``robust'' FF moment 
for the extraction of $\lqcd$ is the peak position $\xi_{\rm max}$ which (i) has the simplest theoretical
expression for its NMLLA+NLO* energy evolution, and (ii) it is quite insensitive to most of the uncertainties
associated with the method (DG fit, finite-mass corrections, and energy-evolution fit). The hadron
multiplicities measured in DIS jets appear somewhat smaller (especially at high-energy) than those measured 
in $\epem$ collisions~\cite{NMLLA_NLO}, a fact pointing likely to limitations of measuring the FF
only in half (current Breit) $e^\pm$p hemisphere and/or of the determination of the relevant Q scale. The fitted HBP
widths appear to show a larger scatter than observed in $\epem$, which is not unexpected as not all the
measurements used exactly the same Eq.~(\ref{eq:DG}) for the DG fit. 
The skewness 
has the largest uncertainties of all moments.
The errors quoted for the individual $\lqcd$ values include only uncertainties from
the energy fit procedure, but the propagated $\alphasmZ$ uncertainties have been enlarged by a common factor such
that their final weighted average has a $\chi^2/$ndf close to unity. Such a ``$\chi^2$ averaging''
method~\cite{PDG} takes into account in a well defined manner any correlations among the 
4 extractions of $\alphas$, as well as extra systematic uncertainties (such as e.g. variations of $\meff$
in the DG fits). 

The final value of $\alphasmZ$ obtained from the combined fit of all DIS jet FF moments (last column of Table~\ref{tab:1})
is $\alphasmZ$~=~0.119$\pm$0.010, which is consistent (although with four times larger uncertainties) with our
previous value obtained from $\epem$ data ($\alphasmZ$~=~0.1195~$\pm$~0.0022) as well as with the
world-average ($\alphasmZ$~=~0.1185$\pm$0.0006). Work is in progress to combine the $\epem$ and $e^\pm$p FF
data into a common global fit, as well as to determine the  theoretical scale uncertainty~\cite{DdERPR}. The
methodology presented here provides a new promising approach 
for the determination of the QCD coupling constant at NLO accuracy complementary to other existing jet-based
methods --such as jet shapes, and/or on ratios of N-jet production cross sections in $\epem$ and p-p
collisions-- with a totally different set of experimental and theoretical uncertainties.


\end{document}